\documentclass[aps,pra,showpacs,twoside,twocolumn,longbibliography,10pt]{revtex4-1}
\usepackage[colorlinks=true, citecolor=red, urlcolor=blue ]{hyperref}
\usepackage{epsfig,newlfont,amssymb,amsfonts,amsmath,bm,subfigure,palatino,mathtools,amsthm,braket,times,soul,enumitem,color}
\usepackage[normalem]{ulem}
\newcommand{\stkout}[1]{\ifmmode\text{\sout{\ensuremath{#1}}}\else\sout{#1}\fi}
\usepackage[english]{babel}
\usepackage[utf8]{inputenc}
\usepackage{array}
\usepackage{xcolor}
\usepackage{graphics}

\newcommand{\ketbra}[2]{|#1\rangle \langle #2|}
\def\Tr{\text{Tr}}

\usepackage{verbatim}
\usepackage{bbm}
\usepackage{wrapfig}

\usepackage{hyphenat}

\usepackage{orcidlink}

\newlength\figureheight 
\newlength\figurewidth

\begin{document}

\title{Resource generation and dynamical complexities in open random quantum circuits}
\author{
Paranjoy Chaki$^{1,2}$\,\orcidlink{0009-0000-5693-5516}}
\thanks{These authors contributed equally to this work.}

\author{Arkaprava Sil$^{3}$\,\orcidlink{0009-0002-5747-5085}}
\thanks{These authors contributed equally to this work.}

\author{Priya Ghosh$^{1,2}$\, \orcidlink{0009-0000-5908-9407}}
\author{Ujjwal Sen$^{1,2}$\,\orcidlink{0000-0002-0091-5847}} 
\author{Sudipto Singha Roy$^{3}$\, \orcidlink{0000-0002-6603-7661}
}

\affiliation{
$^{1}$Harish-Chandra Research Institute,  
Chhatnag Road, Jhunsi, Prayagraj 211019, India\\
\(^2\) Homi Bhabha National Institute,  Training School Complex, Anushakti Nagar, Mumbai 400 094, India\\
$^{3}$Department of Physics, Indian Institute of Technology (ISM) Dhanbad, Dhanbad 826004, India
}

\begin{abstract}

Realistic quantum devices are inherently open and often involve environments with memory. Here, we investigate quantum resource generation in two classes of random circuits, namely, memoryless open and memoryful open random circuits, and compare their behavior with the well-explored random unitary circuit model.  
We show that environmental memory qualitatively alters the dynamics: while unitary and memoryful circuits exhibit sustained growth and saturation of entanglement and non-stabilizerness (magic); memoryless dynamics leads to a distinct behavior where entanglement decays to zero after transient growth, even though non-stabilizerness
remains non-zero, indicating the persistence of nonclassical features beyond entanglement. 
Interestingly, within the  system size and timescale considered in our work, we observe that memoryful circuits more effectively approach low-order quantum-state $k$-designs than the other two circuits. Finally, we show that the Krylov complexity reveals suppressed spreading of quantum states in memoryless circuits, in contrast to strong growth in unitary and memoryful dynamics, which saturates at the maximum value.  
Closed dynamics are therefore usually the most resource-generating, but are ideal; realistic dynamics are open and seem to generate less, but if they possess memory, they can sometimes even outdo closed dynamics.



\end{abstract}

\maketitle
\section{Introduction}
Random quantum circuits have become a powerful and versatile framework for modeling generic quantum dynamics in many-body systems. Constructed from layers of local random unitary gates, these circuits offer a minimal yet effective setting for investigating the nonequilibrium behaviors of quantum systems.
 Despite being constructed from random gates, these circuits capture universal features of quantum dynamics~\cite{sagar_vijay_ruc}, with critical exponents that accurately match those predicted by the Kardar–Parisi–Zhang (KPZ) equation ~\cite{KPZ}, revealing a deep connection between quantum information spread and classical stochastic growth processes. In addition, random unitary circuits have also been used to study the growth of bipartite~\cite{sagar_vijay_ruc,RUC_1} and multipartite entanglement as quantified through the geometric measure of entanglement~\cite{GGM_RUC}.  At the same time,  spreading of other quantum resources such as quantum magic~\cite{Magic_d},  non-Gaussianity ~\cite{Resource_spreading}, and quantum chaotic properties~\cite{holevo,Q_scrambling,Aranya@2024, Suchsland@2025} have been systematically investigated within this framework, together with its role in the context of deep thermalization~\cite{Deep_t}. See more ~\cite{Nahum_MIPT,Rand_Mpemba1,Mpemba_2,mpemba_3,Mpemba_4}.


Despite covering such an extensive range of phenomena, most  prior works have focused on closed-system dynamics generated by random unitary gates. However, realistic quantum devices are inherently open: qubits interact with environments. Understanding how these interactions affect the evolution of quantum resources is crucial both for fundamental studies of quantum dynamics and for practical applications in quantum technologies.

In this work, we investigate how the dynamics of random quantum circuits are modified in open quantum systems and explicitly study the role of memory effects. We show that memory  acts as effective control parameters governing the generation of quantum resources, dynamical complexity, and randomness of the resulting dynamics. 
To this end, we consider three classes of random quantum circuits. The first is a closed-system random unitary circuit (RUC), already studied extensively in the literature~\cite{RUC_1,RUC_2,RUC_3,RUC_4,RUC_5,RUC_6,RUC_7,GGM_RUC}, in which two-qubit random unitary gates are arranged in a brick-wall pattern with periodic boundary conditions. In realistic settings, however, it is difficult to keep a system completely isolated from its environment. When the system interacts with external auxiliary degrees of freedom, its dynamics become open. 
Incorporating such system–auxiliary interactions within our random circuit framework naturally leads to two classes of open random circuits. The first consists of memoryless open random circuits (MLORC), where at each time step a fresh, randomly prepared auxiliary interacts with the system. The second comprises memoryful open random circuits (MFORC), where a randomly chosen auxiliary qubit is introduced initially and then retained throughout the dynamics, thereby introducing memory effects. 
\begin{figure}[h!]
    \centering
    \includegraphics[width=\linewidth]{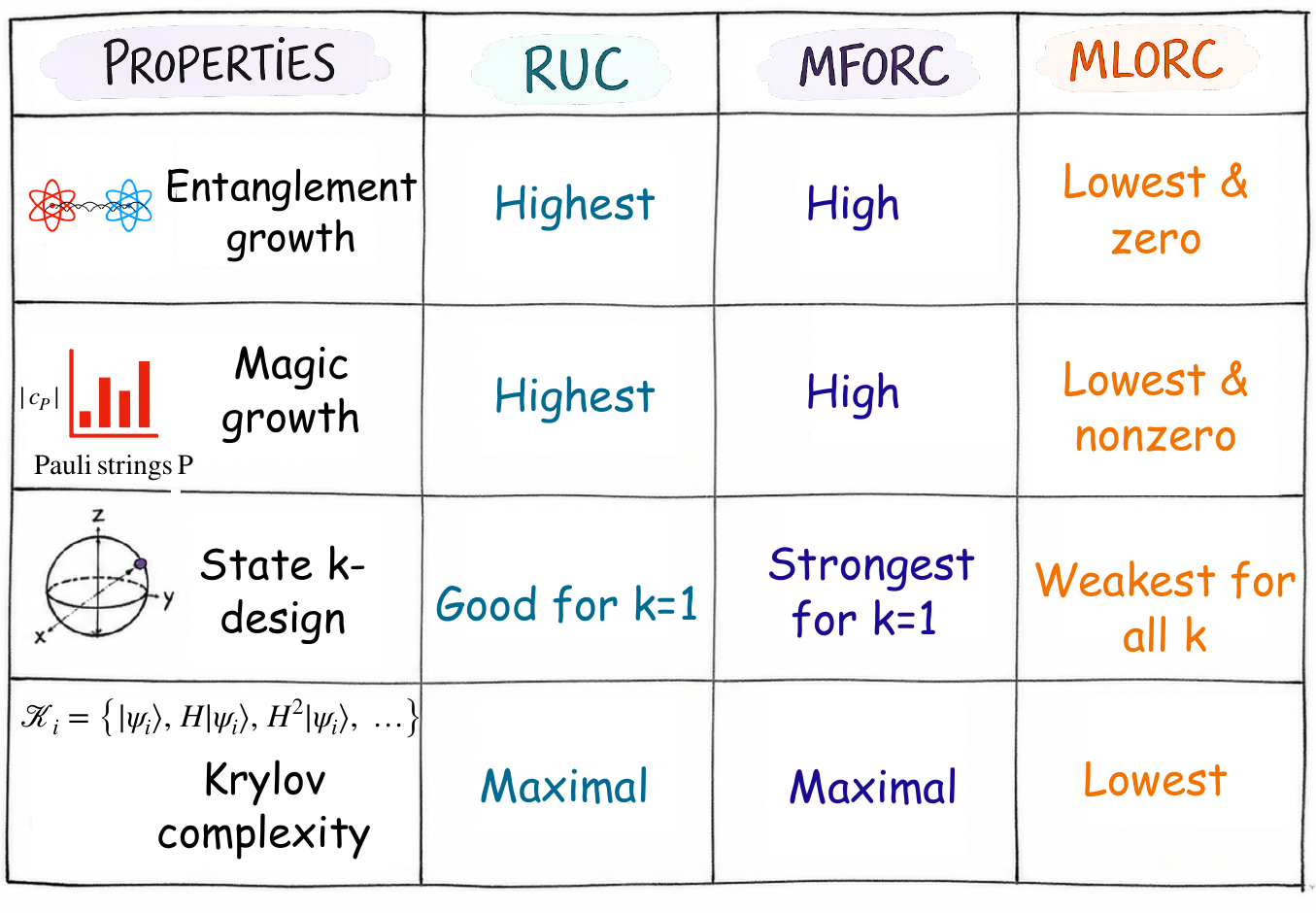}
    \caption{Comparison of different classes of random quantum circuits
    in terms of resource generation power and complexity measures within the system sizes and timescales accessible in our simulations}.
    \label{Schematic1}
\end{figure}

To systematically probe these effects, we adopt two complementary perspectives: (i) a resource-based viewpoint or quantum state preparation complexity, by focusing on entanglement and non-stabilizerness, and (ii) a dynamical and statistical viewpoint, capturing randomness and dynamical complexity  via quantum state $k$-designs and Krylov complexity, respectively. These perspectives allow us to understand how memory influence both the generated quantum resources and the nature of the underlying dynamics.

As a starting point, we first investigate the behavior of logarithmic negativity (log-neg)~\cite{L_1,L_2,L_3}, which is a well suited measure of entanglement for both pure and mixed states for all three circuits mentioned above.
 In both RUC and MFORC, the entanglement initially grows from a vanishing value and eventually saturates at long times. Notably, although the saturation value attained in the random unitary circuit is slightly higher than that in open random circuit with memory, the overall behavior remains qualitatively similar, indicating that  memory effects help preserve entanglement structure even in open dynamics.  In contrast, for memoryless open random circuits, the entanglement increases only at the first few time steps and then rapidly decays, ultimately saturating to zero.
 These features already emerge in small system sizes, making them accessible to near-term quantum devices.

To complement this behavior, we further study quantum state preparation complexity by examining non-stabilizerness (magic), which reflects both the difficulty of classical simulation and the presence of genuinely quantum computational resources. We quantify non-stabilizerness using the second Rényi stabilizer entropy (SRE)~\cite{SRE} for all three classes of circuits. We find that the growth of magic closely mirrors the behavior of entanglement qualitatively. For both RUC and MFORC, the SRE increases with circuit depth and eventually saturates, indicating sustained generation of non-stabilizer resources. In contrast, for memoryless open random circuits (MLORC), the SRE saturates at a significantly lower value, reflecting the suppressing effect of environmental decoherence. Interestingly, despite suppressed entanglement, MLORC can still generate separable yet non-stabilizer states. These results highlight the crucial role of memory in sustaining both entanglement and non-stabilizer resources. 

\begin{figure*}
    \centering
    \includegraphics[width=0.85\textwidth, keepaspectratio]{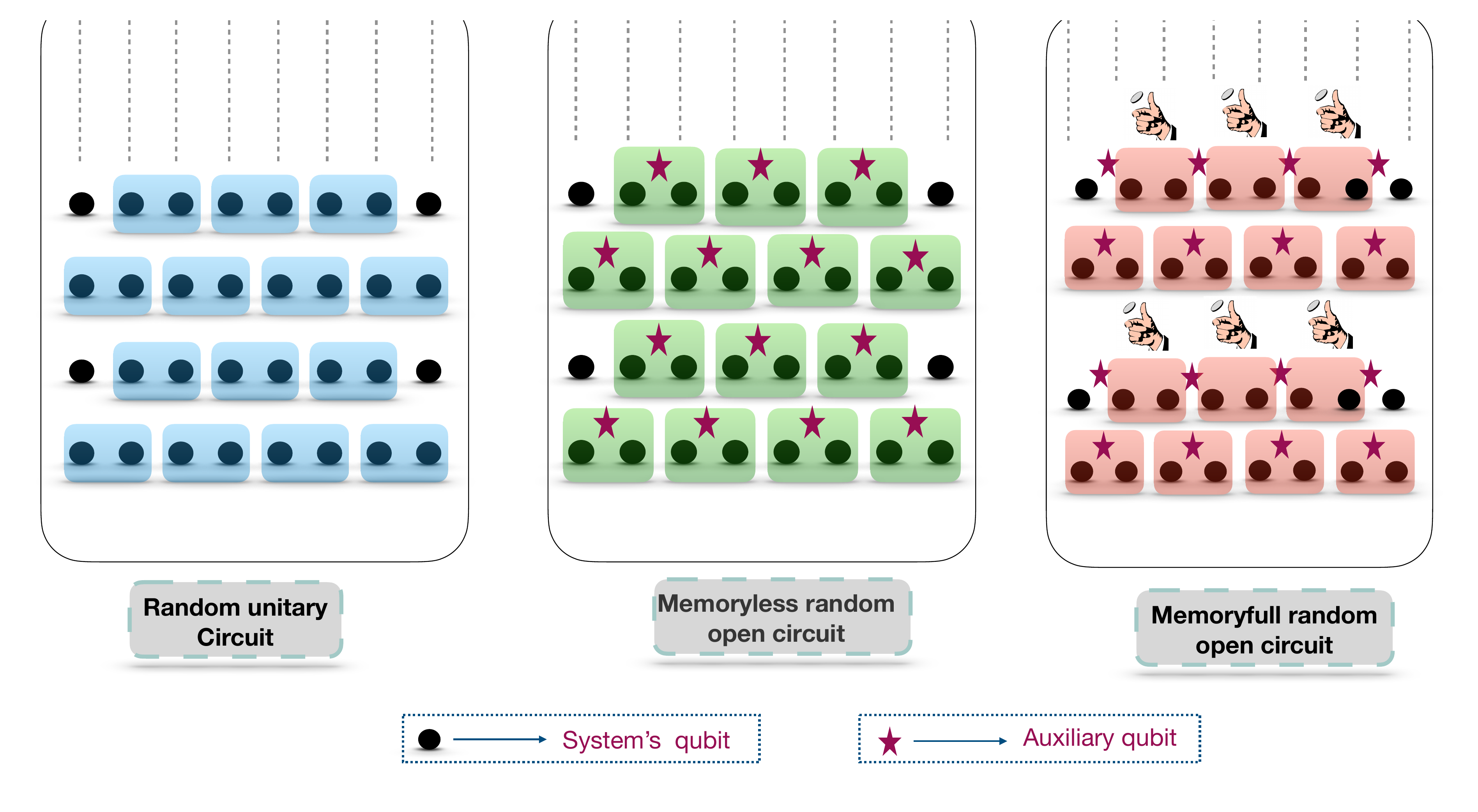}
    \caption{\textbf{Schematic diagram of closed and open   random quantum circuit. } In all three schematics, the black dots denote the qubits comprising the respective circuits. The left panel represents a unitary random circuit. In contrast, the middle panel illustrates a memoryless open random circuit, where no memory effects arise since the auxiliary qubits are discarded after each time step and replaced by fresh ones (sampled randomly). Finally, the rightmost panel depicts a memoryful open random circuit, in which the auxiliary qubits (sampled initially) are retained throughout the dynamics, thereby inducing memory effects. For all the random circuits, in the initial step, the map acts on the {$\{1,2 \},\{3,4\},\ldots,\{2n-1,2n\}$} pairs of qubits. In the subsequent step, the maps act on the $\{2,3\},\{4,5\},\ldots,\{2n-2,2n-1\}, \{2n,1\}$ pairs. This process continues in the same manner for all three random circuits. The only exception is that, for memoryful open random circuit, in the even time steps, the choice of auxiliary system is determined by a coin toss.}
    \label{amp-w-time_1}
\end{figure*}

In the second part of our analysis, we first assess the emergence of genuine randomness in the states generated by these circuits using quantum state $k$-designs~\cite{Unitary-t_1, Unitary-t_2, Unitary-t_3, Unitary-t_4, Unitary-t_6, unitary-t_main, Unitary-t_7, Unitary-t_8, Unitary-t_9, Unitary-t_10, Unitary-t_11}, which test whether the ensemble reproduces Haar-random statistics up to the $k$-th moment. This provides a more stringent and higher-order benchmark beyond entanglement, and non-stabilizerness, by capturing correlations across multiple moments of the state ensemble.   In contrast to the trends observed for other quantities, we find that within the system sizes and timescales accessible in our simulations, the presence of memory enhances pseudorandomness, enabling more efficient approximation of low-order $k$-designs compared to even random unitary circuit dynamics. 

Finally, we characterize the dynamical complexity of circuits, which originates from their quantum chaotic properties, using the Krylov complexity measure~\cite{Parker@2019, Caputa@2022, Aranya@2022_KrylovOpen1, Aranya@2023_KrylovOpen2, Liu@2023, Caputa@2024, Garrido@2024_thesis, Budhaditya@2025, Pratik@2025_review, Rabinovici@2025_review}.  Krylov complexity has emerged as a powerful diagnostic of quantum chaos~\cite{Balasubramanian@2022, Rabinovici@2022, Hashimoto@2023, Scialchi@2024, Balasubramanian@2025, Alishahiha@2025, Baggioli@2025}. While random unitary circuits (RUC) are known to exhibit strong chaotic behavior, we find that this behavior persists even when the dynamics are extended to open quantum settings with memory (MFORC): both random unitary circuits and  open random circuits with memory exhibit strongly chaotic behavior as in both the cases, the dimension of the Krylov space grows rapidly and saturates close to its theoretical upper bound. In contrast, for memoryless open random circuits, the Krylov space dimension remains significantly below this bound, indicating reduced chaoticity.  We stress here that the previous studies of Krylov complexity in open quantum systems primarily focus on continuous-time dynamics generated by Lindblad Liouvillians~\cite{Aranya@2022_KrylovOpen1, Aranya@2023_KrylovOpen2, Liu@2023, Pratik@2025_review, Rabinovici@2025_review}. In contrast, our work investigates the growth of Krylov complexity in random quantum circuit architectures, where the evolution is governed by discrete layers of random quantum channels.

Overall, our results show that environmental memory qualitatively reshapes the growth of quantum resources, non-stabilizerness, dynamical complexity, and pseudorandomness in random quantum circuits. Rather than acting solely as a source of decoherence, system–environment interactions can selectively preserve entanglement, non-stabilizerness, and dynamical complexity, while simultaneously enhancing low-order randomness. These findings highlight the nontrivial role of memory effects in open quantum dynamics and suggest new possibilities for controlling quantum resource generation in realistic noisy quantum devices. Schematic presented in
Fig.~\ref{Schematic1} summarizes the main things to be discussed in the main
text in detail.

The manuscript is organized as follows. In Sec.~\ref{2s}, we provide a brief overview of the three classes of circuits considered in this work. In Sec.~\ref{3s}, we present the dynamical behavior of entanglement, followed by an analysis of the growth of magic across all three circuit  in Sec.~\ref{4s}.  In Sec.~\ref{sec:k:design}, we discuss the emergence of quantum state $k$-designs, and in Sec.~\ref{5s}, we investigate Krylov complexity as a measure of dynamical complexity for these circuits. Finally, we summarize our findings and conclude in Sec.~\ref{6s}.

\section{Setting the stage}\label{2s}
 We start our discussion by introducing the variety of quantum circuits we have considered in our work, namely, the random unitary circuit, the memoryless open random circuit, and the memoryful open random circuit. Below we briefly introduce them.

 {\bf \it Random unitary circuit (RUC):} 
The first variant of random circuit that we consider in our work is the random unitary (RU)~\cite{RUC_1,RUC_2,RUC_3,RUC_4,RUC_5,RUC_6,RUC_7,GGM_RUC} circuit composed of two-qubit random unitary gates, as illustrated in the leftmost panel of Fig.~\ref{amp-w-time_1}. Random unitary gates ($U_i$) are represented by blue boxes, while qubits are shown as solid black circles.  At every time step, such  $U_i$'s are drawn {Haar-uniformly} from the unitary group $\mathrm{U}(4)$.
The execution of the circuit on an initial state can be realized as follows: At the first time step, two-qubit random unitary gates $U_i$ are applied on pairs $(2i-1, 2i)$. In the next step, the unitaries $U_i$ act on the pairs $(2i, 2i+1)$. Here $i$ denotes a non-zero integer. We additionally employ periodic boundary conditions by coupling the first and the $N$th qubits through an additional global two-qubit unitary. 
While computing the relevant physical quantities, we perform averaging over different realizations generated from randomly sampled unitaries and initial states.

 \textbf{\it Memoryless open random circuit (MLORC):} 
The second example of a random circuit goes beyond the unitary assumption. Here, we consider a more general scenario by considering the interaction between the system and the external auxiliaries. 
The schematic representation of this random circuit is shown in the middle panel of Fig.~\ref{amp-w-time_1}. The qubits of the main system are represented by solid black circles, the random unitary gates by green boxes, and the auxiliary systems by red stars. The evolution of the initial state can be realized through the following steps: At the first time step, a three-qubit random unitary $\mathcal{U}_i$ is applied to a two-qubit pair $(2i-1, 2i)$ (where $i$ is a non-zero integer), together with an auxiliary qubit, sampled randomly (all of them chosen Haar-uniformly). In the next time step, three qubit unitaries are sampled again uniformly and act on a two-qubit pair $(2i, 2i+1)$ along with an auxiliary qubit. It should be noted that the random unitaries and the auxiliary systems are independently resampled at each time step. In particular, at every step new auxiliary qubits are introduced, and no information from previous steps is retained. As a result, no memory effect is present in the dynamics, and the evolution remains Markovian (memoryless). For this reason, we refer to the circuit as a memoryless random open circuit. Here, we also impose periodic boundary conditions by coupling the first and the $N$th qubits of the system.

 \textbf{\it  Memoryful open random circuit (MFORC):}  The final example we wish to put forward is the open random circuit with memory. Unlike memoryless open random circuits, auxiliary systems are not sampled afresh in each time step; rather, the auxiliary qubits are retained throughout the dynamics and thus memory effect emerges. In the right panel of the schematic diagram, the circuit diagram of open random circuit with memory is depicted, where qubits of the main system are represented by solid black circles, the random unitary gates by pink boxes, and the auxiliary systems by red stars.

At each odd time steps, random three-qubit unitary gates (represented by pink boxes) act on the pair $(2i-1, 2i)$ and its associated auxiliary system. Similarly, at even time steps, the three-qubit unitaries are applied to the pairs $(2i,2i+1)$, together with an auxiliary system. In our protocol, at this stage, there exist two possible choices of auxiliary systems: one with the pair $(2i-1, 2i)$, and another with the pair $(2i+1, 2i)$. The auxiliary system to be used is randomly selected by a coin-tossing procedure: if the result is heads, the auxiliary corresponding to $(2i-1, 2i)$ is chosen; otherwise, the one associated with $(2i+1, 2i)$ is used.  

Therefore, our choice of circuits follows the standard random-circuit framework. The RUC, a brick-wall array of Haar-random two-qubit gates with periodic boundary conditions, serves as the closed-system reference. The open circuits are obtained from it by a minimal modification: each gate acts jointly on a system pair and an auxiliary qubit playing the role of the environment, so that the system evolves under a random quantum channel. This allows a memoryless and a memoryful random circuits to be compared within the same family of random gates.

Reusing the auxiliary qubits (acts as environment) allows the environment to carry information from earlier interactions and mediate correlations between different parts of the system. However, the present setup does not allow us to distinguish the individual contributions of memory and environmental correlations to entanglement preservation. Therefore, the differences between MLORC and MFORC should be attributed to their combined effects rather than to memory alone. Separating their respective roles would require a further study in which memory effects and environmental correlations are controlled independently.

Additionally, we use Haar-random three-qubit unitaries to model the system–environment interaction without introducing an explicit coupling parameter. Different random unitaries can correspond to different effective interaction strengths, from weak to mederate to strong. In our work, however, we average the behaviour over many such random configurations and find that the qualitative behaviour remains the same. Studying the transition from weak to strong coupling using an explicit coupling parameter is an interesting direction, but is beyond the scope of the present work.

\section{Quantum resource growth}
One of the main objectives of our work is to examine how quantum resources are generated in different kinds of random quantum circuits we introduced above and how distinct or close they remain to each other. Towards that aim, below  we first  analyze the growth of entanglement and mutual information of all three types of random circuits considered in our study. Thereafter, we present the results of growth of magic (or non-stabilizerness) as another resource. Since both entanglement and non-stabilizerness are closely connected to nonclassical correlations and the classical simulability of quantum states, they together provide important insights into the complexity of quantum state preparation. Hence, throughout this work we refer to them collectively as measures of quantum resource generation and state-preparation complexity.

\subsection{Entanglement dynamics in random quantum circuits
}\label{3s}


\begin{figure}[h!]
    \centering
    \includegraphics[width=1.0\linewidth]{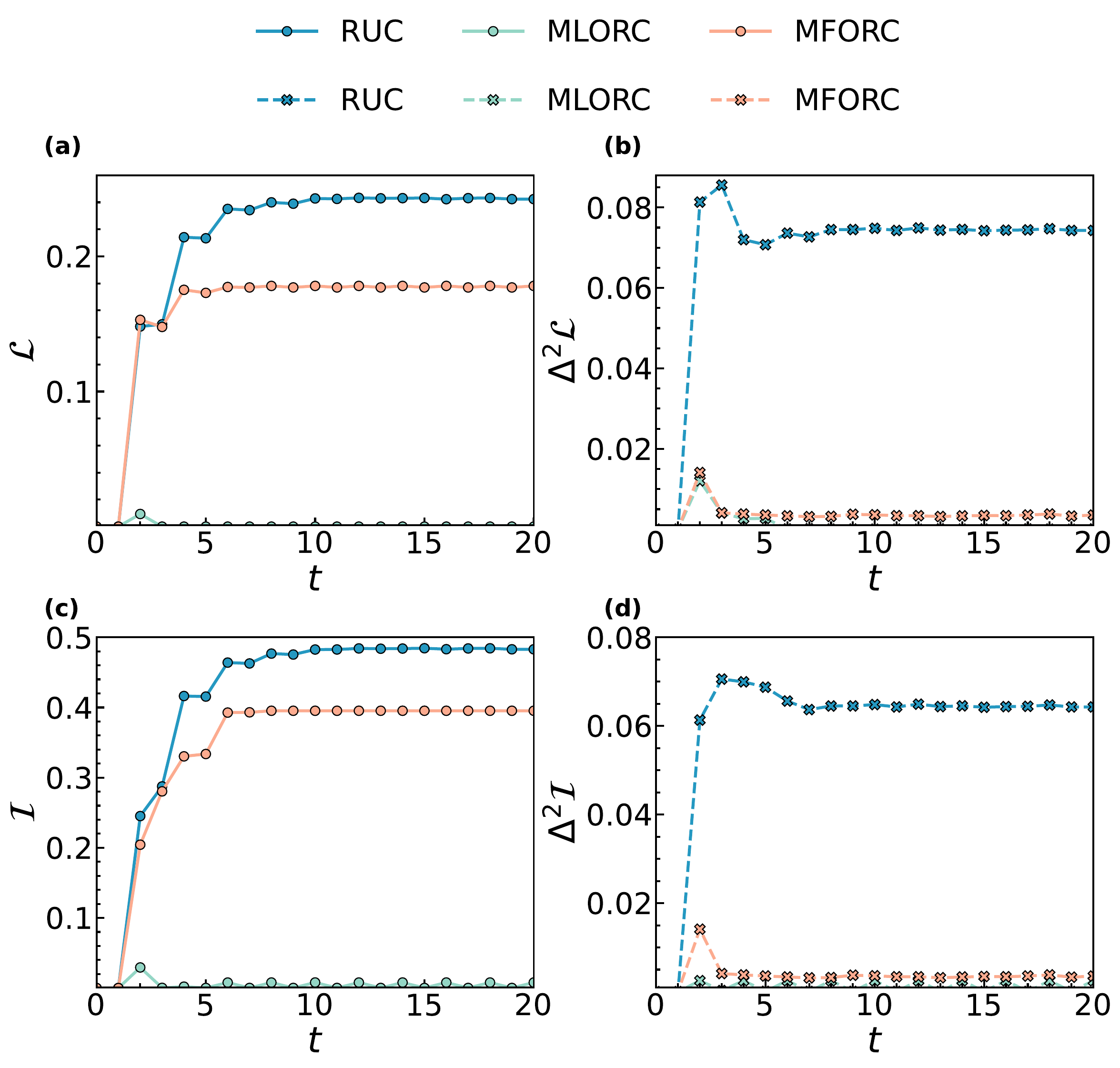}
    \caption{\textbf{Growth of entanglement and quantum mutual information with time step.}  (a) Describes the growth of entanglement (log-neg) for unitary, memoryless open random circuits, and memoryful open random circuits. The vertical and horizontal axes denote the log-neg ($\mathcal{L}$) as the measure of entanglement and the time step $t$, respectively. The curves corresponding to unitary, MLORC, and MFORC are denoted by sky blue,  green, and orange  colors, respectively. (c) Shows similar behavior obtained for mutual information  ($\mathcal{I}$). We additionally plot the fluctuation arising in both entanglement $\Delta^2\mathcal{L}_N$ and mutual information $\Delta^2\mathcal{I}$ for all three random circuits in (b) and (d), respectively. Here, 
the total number of qubits, including both system ($N_s=8$) and auxiliary degrees of freedom ($N_a=4$) is $12$.}
    \label{f2x}
\end{figure}

Entanglement~\cite{L_2} is a vital resource that offers significant advantages in a broad range of quantum information processing tasks, including quantum teleportation~\cite{Teleportation}, quantum dense coding~\cite{dens_coding}, secure communication protocols such as quantum cryptography~\cite{crypto}, and quantum metrology~\cite{metro}.
Although several measures of entanglement exist in the literature, we stick to the logarithmic negativity (log-neg)~\cite{L_1,L_2,L_3} as our chosen measure of entanglement.
Let us consider a bipartite system in a quantum state $\rho_{AB}$ acting  on the Hilbert space $\mathcal{H}_A\otimes \mathcal{H}_B$. Now if we consider partial transposition of the state with respect to the party $B$, $\rho^{T_B}_{AB}$, the measure negativity is defined as $\mathcal{N}=\sum_i|\lambda_i|$. Here, $\lambda_i$ are the negative eigenvalues, i.e., $\lambda_i<0$ of $\rho^{T_B}_{AB}$. Hence, for a separable state $\mathcal{N}$ is zero, and any non-zero value quantifies the amount of entanglement the quantum state has in the bipartition, $A:B$. The logarithmic negativity (log-neg) thus can be  defined as 
\begin{equation}
\mathcal{L}=\log_2(2\mathcal{N}+1).
\end{equation}
Likewise, the mutual information~\cite{N_C} of the state $\rho_{AB}$ in the bipartition $A:B$ can be mathematically expressed as
\begin{equation}
  \mathcal{I}(A:B)=S(\rho_A)+S(\rho_B)-S(\rho_{AB}).
\end{equation}
Here, $S(\rho_A)$, $S(\rho_B)$, and $S(\rho_{AB})$ are the von Neumann entropies of the subsystems $A$, $B$, and $AB$ in states $\rho_A$, $\rho_B$, and $\rho_{AB}$, respectively. From now on, instead of $\mathcal{I}(A:B)$, we will use the symbol  $\mathcal{I}$ to denote the mutual information between the bipartition $A:B$. Entanglement can depend significantly on the choice of bipartition. In all plots showing entanglement dynamics, we explicitly specify the bipartitions considered.

We present the behavior of entanglement and mutual information for all the considered circuits in Fig.~\ref{f2x}. 
The initial state of all three types of random circuits is considered as
\begin{equation}\label{1}
\rho_{\mathrm{in}} = \bigotimes_{i=1}^{N/2} \rho^{i}, 
\end{equation}
where $\rho^{i}$ are two-qubit density matrices corresponding to qubit pairs $(2i-1, 2i)$ sampled Haar uniformly. The auxiliary systems are randomly sampled at each time step in the case of MLORC, whereas for MFORC the initially sampled auxiliary is preserved throughout the entire dynamical evolution, as described in Sec.~\ref{2s}. We consider a system of size $N_s=8$ and compute the logarithmic negativity in the $N_s/2:N_s/2$ bipartition (i.e., $1234:5678$). From Eq.~\eqref{1}, It can be seen that the initial state remains product in that bipartition.   Starting from this zero-entanglement state, we investigate which of the considered circuits is more effective in generating entanglement during the dynamics.
In Fig.~\ref{f2x}(a), the variation of logarithmic negativity with respect to time step $t$ for RUC, MLORC and MFORC are shown by the sky blue, green, and orange smooth curves with markers, respectively. The entanglement is quantified in ebits, and time is treated as a dimensionless quantity throughout this work. The number of random realizations, taken over both random initial states and unitaries, is $500$.
We observe that the  behavior of logarithmic negativity for random unitary and open random circuits with memory are close, exhibiting initial growth with iteration time 
$t$ and eventually saturating to the values 
$ 0.24$ and $ 0.18$, respectively, with some numerical precision. This implies that although the saturation of the entanglement for an open random circuit with memory occurs earlier than that of the unitary circuit, its saturation value is lower than that of the random unitary circuit.

\begin{figure}[h!]
    \centering
    \includegraphics[width=0.8\linewidth]{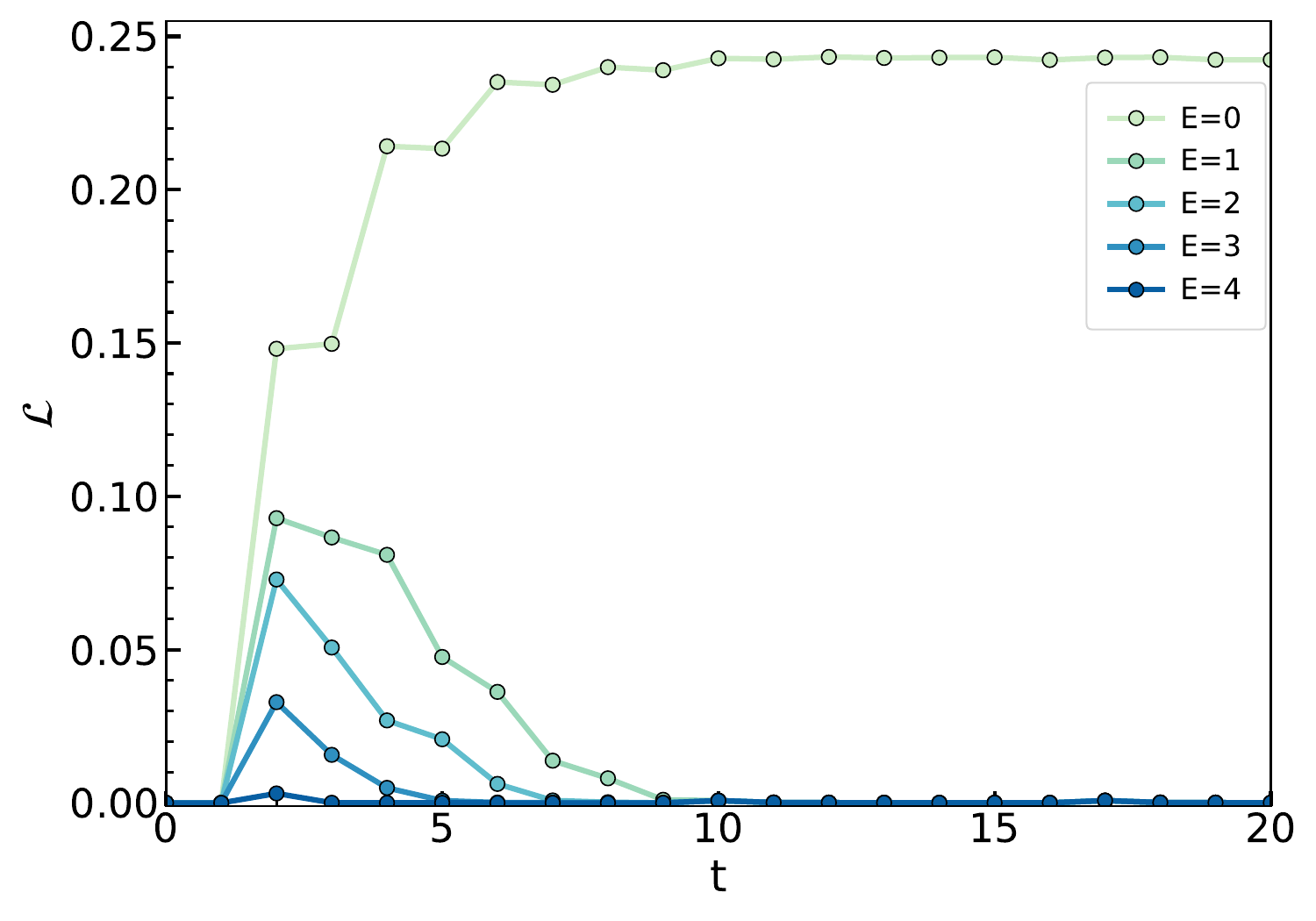}
    \caption{\textbf{Robustness of growth of entanglement under the action of MLORC.}  We present the behavior of the dynamical evolution of the entanglement for different numbers of auxiliary systems, $E=0,1,2,3,4$. The corresponding curves are distinguished by sequential colors, as indicated in the legend. The system size is fixed at $N_s=8$.}
    \label{rcc_scaling}
\end{figure}

In contrast, the behavior of the entanglement obtained for the MLORC remains drastically different from that of the RUC and MFORC. Here we observe that the action of memoryless open random circuit suppresses the growth of entanglement completely. The circuit loses its entanglement just after two iterations. The behavior of the mutual information in this case remains qualitatively similar, as shown in Fig.~\ref{f2x}(c).

 As we have performed  averaging over the random realizations, there exists a fluctuation about the mean value. The fluctuation of any general observed quantity $O$ is defined as
\begin{equation}
    \Delta^2 O = \langle O^2 \rangle - \langle O \rangle^2 .
\end{equation}
In our work, we calculate numerically the fluctuation in entanglement $\Delta^2\mathcal{L}$ and mutual information $\Delta^2\mathcal{I}$ that arise in each time step $t$  due to the random realization consideration. In Fig.~\ref{f2x}(b) and Fig.~\ref{f2x}(d), we plot the fluctuations $\Delta^2\mathcal{L}$ and $\Delta^2\mathcal{I}$ about the mean values $\mathcal{L}$ and $\mathcal{I}$ respectively. From both the figures, it is visible that for unitary circuits there exists an appreciable amount of fluctuations corresponding to both the quantities $\mathcal{L}$ and $\mathcal{I}$, whereas the fluctuation almost vanishes for MLORC and MFORC. However, we also check other bipartitions and find the same qualitative behavior for both contiguous and non-contiguous partitions. Additionally, due to periodic boundary conditions, equivalent translations (e. g. [$1234:5678$], [$2345:6781$]) of a given bipartition yield the same results.

Additionally, we consider the case where limited number of auxiliary qubits interacts with the system qubits during the dynamics in memoryless random circuit. The result is shown in Fig~\ref{rcc_scaling}. We observe that even if a single auxiliary interacts with the system, suppression of entanglement begins at the second iteration time and the logarithmic negativity rapidly decreases thereafter. In this case, the entanglement survives up to about eight iterations before it vanishes. As the number of  interacting  auxiliaries increases, the entanglement decays more rapidly. For example, as shown in Fig.~\ref{rcc_scaling}, when  two pairs of auxiliaries interact with the  system qubits, the logarithmic negativity survives only up to approximately six iterations before going to zero, and so on. Hence, the near monotonic reduction of entanglement lifetime with an increasing number of auxiliary qubits suggests that  continual injection of fresh auxiliary degrees of freedom in MLORC,  effectively acts as a source of decoherence and prevents sustained entanglement buildup.

A careful analysis of the reduced quantum states of the subsystems provides a clear explanation for this behavior. We show that for an MLORC, the long-time composite state of system qubits and auxiliaries qubit behaves similarly to the generalized GHZ states. As a consequence, after tracing out the auxiliary qubit (or any system qubit), the entanglement in any bipartition of the system vanishes.
In contrast, for open random circuits with memory, the long-time state effectively exhibits W-state-like characteristics. Therefore, even after tracing out the auxiliary qubit, the entanglement in any bipartition of the system remains non-vanishing. In Appendix~\ref{app_1}, we provide a detailed discussion of this analysis.

\subsection{Non-stabilizerness generation in  random circuits}\label{4s}

Non-stabilizerness (magic) quantifies how far a quantum state deviates from the stabilizer manifold, and thus captures its degree of classical intractability. Importantly, it constitutes a resource distinct from entanglement: while entanglement characterizes nonlocal correlations, non-stabilizerness determines whether a state can be efficiently simulated by stabilizer-based classical algorithms, irrespective of its entanglement content.
Mathematically, non-stabilizerness can be quantified in various ways, such as the robustness of magic~\cite{R_SRE1,R_SRE_2,R_SRE}, stabilizer Renyi entropy (SRE)~\cite{SRE}, or mana~\cite{mana,Mana_2}. In our work, we stick to  stabilizer Renyi entropy (SRE) as the measure of non-stabilizerness. Let us consider $\mathcal{P}_N$ to be the set of all the Pauli strings of $N$ qubits. Then for any pure multiqubit state $\ket{\psi_N}$, let us define the overlap $\mathcal{E}_P=\bra{\psi_N}P\ket{\psi_N}$, where $P\in\mathcal{P}_N$ and $d=2^N$. The $\alpha$ R{\'e}nyi entropy of the state is given by

\begin{equation}
\tilde{M}(\ket{\psi_N})_{\alpha}=(1-\alpha)^{-1} \Big[\log\sum_{P\in\mathcal{P}_N}|\mathcal{E}_P|^{2\alpha}/d\Big]-\log{d}.
\label{sre:original}
\end{equation}

In our work, we focus on the stabilizer R{\'e}nyi-2 entropy, generalized to mixed states~\cite{SRE}, which is defined as
\begin{equation}
    M(\rho)=\tilde{M}_2(\rho)-S_2(\rho).
\end{equation}
Here, $S_2(\rho)$ is the R{\'e}nyi-2 entropy expressed as  $S(\rho)=-\log\Tr(\rho^2)$. For pure state, since $S_2(\rho)$ is zero, we get back Eq.~(\ref{sre:original}).

\begin{figure}[h!]
    \centering
    \includegraphics[width=1.0\linewidth]{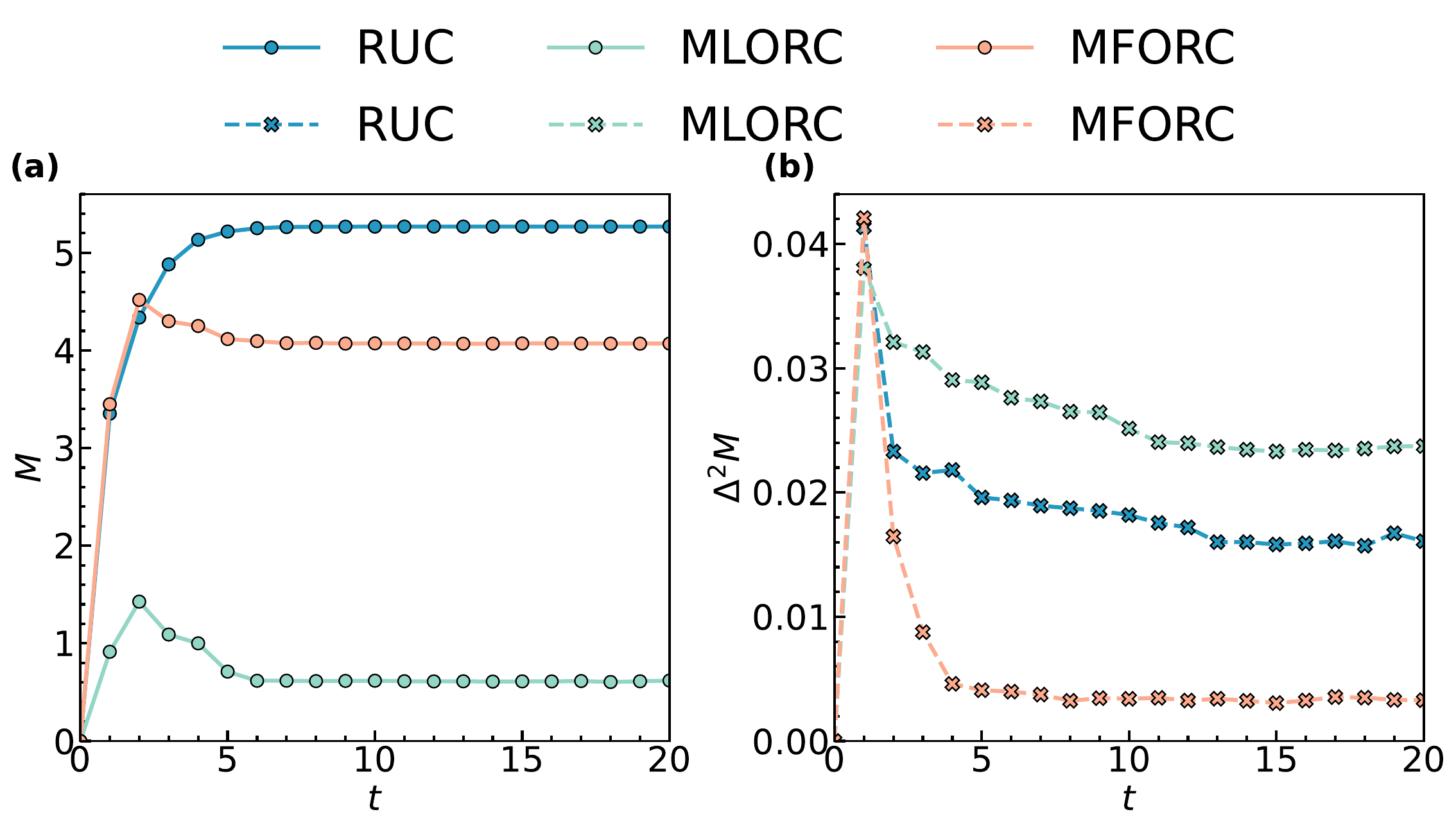}
    \caption{\textbf{Growth of non-stabilizerness as quantified by SRE and the fluctuation associated with it. } Here the vertical and horizontal axes denote the SRE ($M$) and the time step ($t$), respectively. The curves corresponding to unitary, MLORC, and MFORC are denoted by skyblue,  green, and orange colors, respectively. At the same time, fluctuation with respect to the mean value is also denoted by the shade of the same color of the mean curves.}
    \label{f3}
\end{figure}

In Fig.~\ref{f3}, we plot the growth of SRE for random unitary circuits.
Here, we consider the initial states of all three circuits as

\begin{equation}
\rho_{\mathrm{in}} 
= \bigotimes_{i=1}^{N/2} \rho^{i},\hspace{0.4cm} \text{where}\hspace{0.1cm}
\rho^{i}= U_{\mathrm{cli}} \left( \sum_i \mathcal{A}_i M_i \right) U_{\mathrm{cli}}^{\dagger}.\end{equation}

Here, $M_i$ denote the outer products of the two-qubit computational basis states, given by 
$\{\ketbra{00}{00}, \ketbra{10}{10}, \ketbra{01}{01}, \ketbra{11}{11}\}$. The coefficients satisfy $\mathcal{A}_i \geq 0$ and $\sum_i \mathcal{A}_i = 1$, and are chosen randomly. Similarly,   $U_{\mathrm{cli}}$ denotes random Clifford unitaries sampled from the set $\{I, U_H, U_{\frac{\pi}{4}}, U_c\}$. 

\begin{equation}
\scalebox{0.88}{$
U_H = \frac{1}{\sqrt{2}}
\begin{pmatrix}
1 & 1 \\
1 & -1
\end{pmatrix},\quad
U_{\frac{\pi}{4}} =
\begin{pmatrix}
1 & 0 \\
0 & e^{i\pi/4}
\end{pmatrix},\quad
U_c =
\begin{pmatrix}
1 & 0 & 0 & 0 \\
0 & 1 & 0 & 0 \\
0 & 0 & 0 & 1 \\
0 & 0 & 1 & 0
\end{pmatrix}
$}\nonumber.
\end{equation}

Note that when the single-qubit unitaries $U_H$ and $U_{\frac{\pi}{4}}$ are applied to a given two-qubit pair, they are applied to either the first or the second qubit, with the choice made randomly. The auxiliary qubits are sampled in the same manner as described in Sec.~\ref{2s} and Sec.~\ref{3s}.

We consider the initial configuration to be a zero-magic state and investigate the amount of magic generated dynamically in each of the three considered circuits.
Here, to calculate SRE in each time steps we perform the average over 500 random realizations over initial states and unitaries. The system size is $N_s=8$, with an additional $4$ auxiliary qubits, resulting in a total of $12$ qubits. Along the vertical and horizontal axes of Fig.~\ref{f3}, we have plotted the SRE, denoted by $M$, and time, denoted by $t$. We observe that both the unitary and memoryful open random circuits exhibit an initial growth of the SRE with time, followed by saturation to steady-state values. In contrast, the MLORC exhibits a markedly different behavior, with suppressed growth of SRE. Nevertheless, we note here that the non-stabilizerness persists even after entanglement is completely destroyed by the dynamics of MLORC. In other words, MLORC dynamics generates separable but non-stabilizer states. 
The saturation value of the SRE corresponding to the RUC and the MFORC are given by $M^{Sat}_{\mathrm{MFORC}}\approx 4.12$ and $M^{Sat}_{\mathrm{RUC}}\approx 5.21$, respectively, up to the numerical precisions. 

\begin{figure}[h!]
    \centering
    \includegraphics[width=0.8\linewidth]{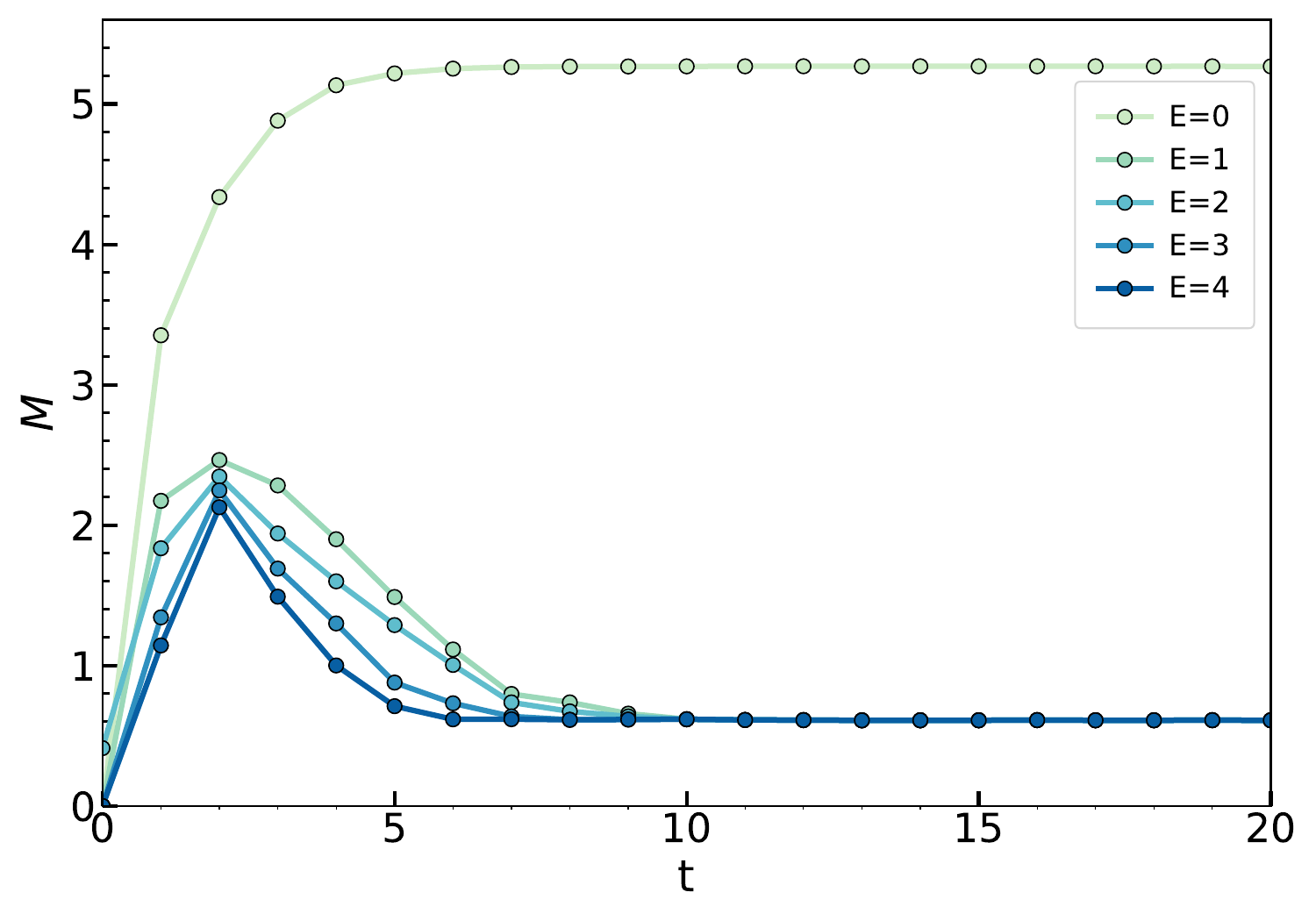}
    \caption{\textbf{Robustness of growth of non-stabilizerness under the action of a random memoryless open random circuit.}
   Here, we plot the dynamical evolution of the SRE for different numbers of auxiliary systems, $E=0,1,2,3,4$. The corresponding curves are distinguished by sequential colors, as indicated in the legend. Here, the system size of the system is considered to be $N_s=8$.} 
    \label{rcc_sre_robustness}
\end{figure}

Although non-stabilizerness poses a challenge for classical simulation of quantum states, it serves as a valuable resource in several quantum information protocols, such as quantum approximation and optimization algorithms (QAQA)~\cite{QAQA}. From this perspective, random unitary circuits are optimal for generating high-magic states; however, maintaining purely unitary dynamics is experimentally demanding due to unavoidable coupling with external environments. Our observations thus show that even in the presence of such external influences, the system can still generate non-stabilizer states, and the presence of environmental memory further enhances this generation. Additionally, similar to entanglement, we examine the robustness of the SRE with respect to the number of system qubit pairs exposed to the environment. However, unlike entanglement, Fig.~\ref{rcc_sre_robustness} shows that increasing the number of auxiliary qubits  influences  both the saturation time and the final saturation value of SRE differently. In particular, unlike entanglement, for smaller qubits the SRE saturates at nearly the same time and reaches almost the same saturation value irrespective of the number of auxiliary qubits considered.

\begin{table}[h]
\centering
\begin{tabular}{|c|ccc|ccc|}
\hline
 & \multicolumn{3}{c|}{\textbf{Entanglement (log-neg)}} 
 & \multicolumn{3}{c|}{\textbf{Non-stabilizerness (SRE)}} \\
\hline
$N_s$ & RUC & MLORC & MFORC & RUC & MLORC & MFORC \\
\hline
4 & 0.11 & $10^{-5}$ & 0.07 & 2.21 & 0.34 & 1.76 \\
\hline
6 & 0.17 & $10^{-6}$ & 0.13 & 3.91 & 0.50 & 3.17 \\
\hline
8 & 0.24 & $10^{-6}$ & 0.18 & 5.41 & 0.61 & 5.28 \\
\hline
\end{tabular}
\caption{Finite-size scaling of the saturation value of entanglement and non-stabilizerness with system size $N_s$, for RUC, MLORC, and MFORC dynamics.}
\label{pop}
\end{table}

Before we move on to the analysis of the dynamical complexity part, here in Table \ref{pop}, we present the finite-size scaling analysis of the saturation values of entanglement and non-stabilizerness obtained for RUC, MLORC, and MORC, for system sizes $N_s=4,6,8$. Here, $N_s$ denotes the system size, but to realize MLORC and MFORC for system size $N_s=8$, we need another $4$ qubit auxiliary system. Therefore, we sample $12$ qubits in total. The table shows that for entanglement, both RU and MFOR circuits show an increment of the saturation value with system size. For MLORC even for smaller system sizes, entanglement follows the trend that already shown in the main figure (Fig.~\ref{f2x}(a)), and remains very low. On the other hand, the SRE increases with the system size for all types of random circuits. For system sizes $N_s=4$ and $N_s=8$, we consider an equal bipartition of the system and evaluate the entanglement between the two halves. In contrast, for $N_s=6$, we compute the entanglement across a $2{:}4$ bipartition, as the system remains initially in a product state for this choice of partition. We emphasize that, although the system sizes considered here are relatively small, the scaling and saturation behaviors of the quantities studied already exhibit clear and consistent trends across the different circuit classes.

\section{Statistical and Dynamical Complexity}
 While state-preparation complexity and resource measures provide important insights into the structure of quantum states, they do not always offer a complete picture of the underlying dynamics or the degree of randomness generated by the circuits. To address this, we now turn to complementary diagnostics that probe the dynamical and statistical aspects of complexity. In particular, we analyze quantum state designs, which quantify how closely the generated ensemble approximates Haar-random states.

\subsection{State $k$-design}
\label{sec:k:design}

\begin{figure*}
    \centering
    \includegraphics[width=0.90\linewidth]{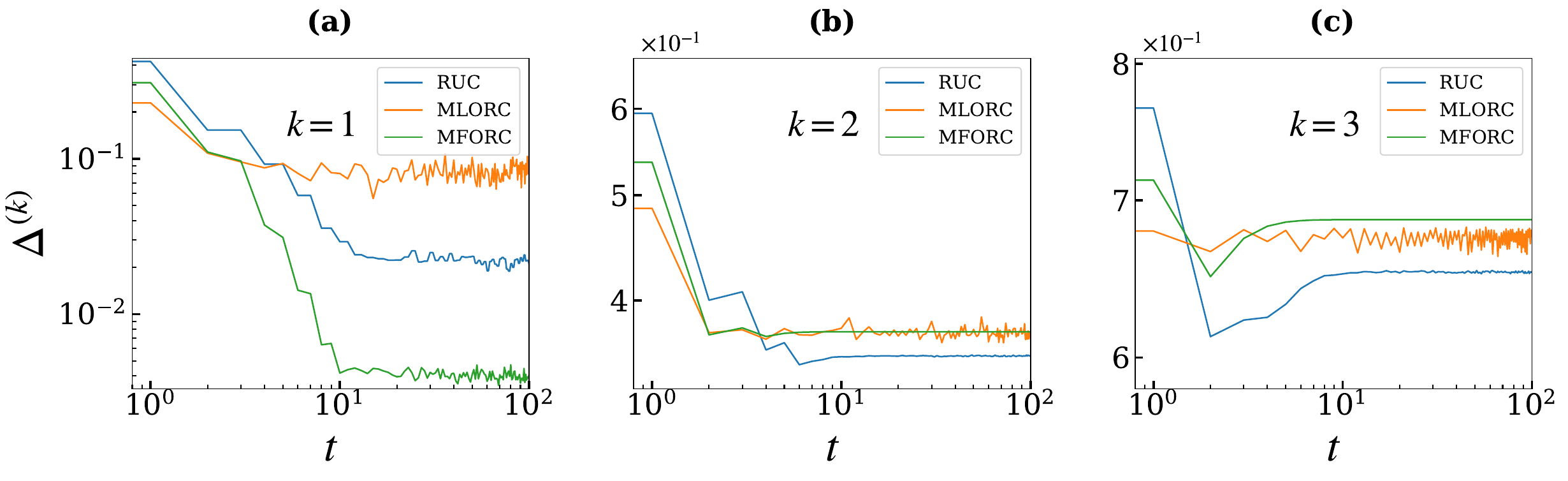}
    \caption{{\bf Emerging state-k design from the random circuits.} Here, we show the time evolution of $\Delta^{(k)}$ for $k=1,2,3$ in panels (a), (b), and (c), respectively, for RUC, MLORC, and MFORC dynamics. The system size is $N_s=8$, with subsystem $A$ consisting of the first two qubits, while projective measurements are performed on the remaining six qubits to construct the projected ensemble. Each curve is averaged over 10 realizations. The behavior of $\Delta^{(k)}$ highlights how different dynamics approach (or fail to approach) Haar-like randomness at increasing moment order.}
    \label{k-design}
\end{figure*}

We begin with the notion of quantum state (k)-designs~\cite{Unitary-t_1, Unitary-t_2, Unitary-t_3, Unitary-t_4, Unitary-t_6, unitary-t_main, Unitary-t_7, Unitary-t_8, Unitary-t_9, Unitary-t_10, Unitary-t_11}, which provide a stringent benchmark for the randomness generated by quantum circuits. An ensemble of states forms an approximate (k)-design if it reproduces the first (k) moments of the Haar measure. Unlike conventional measures that capture only average properties, $k$-designs are sensitive to higher-order correlations and therefore provide a stringent benchmark for randomness. It is well known that random unitary (RU) dynamics can efficiently approximate Haar randomness and thus form approximate $k$-designs at sufficiently long times. Here, we use this as a reference point to benchmark the behavior of MLORC and RU dynamics.

In our approach, we construct the relevant ensemble operationally by performing projective measurements on a subsystem. Specifically, we partition the system into two parts, $A$ and $B$, with $B$ chosen to be larger than $A$. We then perform projective measurements in the computational basis on subsystem $B$. Let $b$ denote a measurement outcome obtained with probability $p_b$, resulting in the post-measurement state $\rho_{AB}^b$. The corresponding state of subsystem $A$ is given by
\begin{align}
    \rho_A^{(b)} = \mathrm{Tr}_B(\rho_{AB}^b).
\end{align}
This procedure generates a projected ensemble $\mathcal{E} = \{p_b, \rho_A^{(b)}\}$.
The $k$-th moment of this ensemble is defined as
\begin{align}
    \rho^{(k)}_{\mathcal{E}} = \sum_b p_b \left(\rho_A^{(b)}\right)^{\otimes k}.
\end{align}
For comparison, the $k$-th moment of the Haar ensemble is given by
\begin{align}
    \rho^{(k)}_{\mathrm{Haar}} = \frac{\Pi_k}{\binom{d + k - 1}{k}},
\end{align}
where $d$ is the dimension of the Hilbert space $\mathcal{H}_A$, and $\Pi_k$ is the projector onto the symmetric subspace of $\mathcal{H}_A$.
The closeness between the projected ensemble and the Haar ensemble is quantified using the trace distance
\begin{align}
    \Delta^{(k)} = \frac{1}{2} \left\| \rho^{(k)}_{\mathcal{E}} - \rho^{(k)}_{\mathrm{Haar}} \right\|_1, 
\end{align}
where $\|A\|_1=\mathrm{Tr}\sqrt{A^\dagger A}$ denotes the trace norm. We say that the ensemble $\mathcal{E}$ forms an $\epsilon$-approximate quantum state $k$-design if $\Delta^{(k)} \leq \epsilon$.

The behavior of $\Delta^{(k)}$ as a function of time is shown in Fig.~\ref{k-design}. For $k=1$, all three dynamics exhibit a clear decay of $\Delta^{(1)}$, indicating convergence toward equilibrium at the level of reduced density matrices, consistent with standard notions of thermalization. However, qualitative differences emerge already at this level, with memoryful  dynamics (MFORC) achieving the smallest steady-state deviation, followed by RUC and memoryless open quantum circuit (MLORC).
However, the differences with the Haar ensemble become significantly more pronounced. While $\Delta^{(k)}$ initially decreases, it quickly saturates to relatively large values, indicating that none of the dynamics fully approximate higher-order Haar randomness within the accessible time scales. {In particular, the persistence of a finite $\Delta^{(k)}$ at late times signals a clear deviation from ideal $k$-design behavior, despite the presence of strong entanglement and dynamical complexity.}

In other words, in an open quantum setting, the formation of approximate quantum state $k$-designs carries a clear imprint of system--environment interactions. In particular, the presence of memory significantly influences the buildup of higher-order correlations in the generated ensemble. Our results show that memoryful open circuits (MFORC) approach low-order $k$-designs more efficiently than both unitary and memoryless dynamics. In contrast, memoryless evolution (MLORC) suppresses the formation of higher-moment randomness. 
These results demonstrate that standard indicators such as entanglement growth and nonstabilizerness do not necessarily imply closeness to Haar-like ensemble, highlighting the more stringent nature of the $k$-design criterion that requires separate analysis. 

 One should note that the closer approach of MFORC to the Haar state ensemble at $k=1$ does not necessarily imply stronger coherent scrambling than in RUC. As discussed in previous studies of noisy quantum circuits~\cite{Harris2022, Yoshida2019} incoherent mixing can also make state statistics appear more Haar-like. For example, if strong noise drives the system towards the maximally mixed state $\rho=I/2^N$, its first moment agrees with that of the Haar ensemble of pure states, even though the state contains no quantum correlations. However, in our simulations, MFORC retains appreciable entanglement and non-stabilizerness, although their saturation values remain below those of RUC. Thus, the closer approach to the Haar benchmark at $k=1$ should be interpreted alongside the entanglement and SRE results.

\subsection{Krylov complexity}
\label{5s}

We now turn to Krylov complexity~\cite{Parker@2019, Caputa@2022, Aranya@2022_KrylovOpen1, Aranya@2023_KrylovOpen2, Liu@2023, Caputa@2024, Garrido@2024_thesis, Budhaditya@2025, Pratik@2025_review, Rabinovici@2025_review}, which provides a complementary characterization of dynamical complexity. While quantum state $k$-designs quantify the emergence of Haar-like randomness in the generated state ensemble, Krylov complexity probes the growth of operators under time evolution.
Essentially, it says how a `simple' initial operator grows along the Krylov chain under time evolution and becomes more `complex' one. The average position of the operator in the Krylov chain gives the measure of Krylov complexity. Here, we briefly review the formalism and then discuss Krylov complexity in our context.\\

\noindent If a density matrix $\rho(0)$ describes the initial state of the system, then solving the  Liouville-von Neumann equation we get the time evolved state, given by ~\cite{Caputa@2024}
\begin{eqnarray}
    \rho(t) = e^{-iHt} \rho(0)\ e^{iHt} = \sum_{n=0}^{\infty} \frac{(-it)^n}{n!}\ \mathcal{L}^n \rho(0),
\end{eqnarray}
where $H$ is the system Hamiltonian and $\mathcal{L}$ is the Liouvillian superoperator whose action on an operator $\mathcal{O}$ is defined by $\mathcal{L}\mathcal{O} = [H, \mathcal{O}]$. Using the operators $\{\rho(0),\  \mathcal{L}\rho(0),\ \mathcal{L}^2\rho(0),\cdots\}$, a set of orthonormal basis $\{|\rho_0),\ |\rho_1), |\rho_2), \cdots\}$ of the Krylov subspace is constructed by Lanczos algorithm, known as krylov basis (here convention wise, we have used $|\ \cdot \ )$ to denote the operators in the operator space). The time-evolved density matrix $\rho(t)$ can be expanded in the Krylov basis using the following form
\begin{eqnarray}
    |\rho(t)) = \sum_{n} i^n \phi_n(t) |\rho_n).
\end{eqnarray}
Here, we use the inner product definition given by $ (A| B) = \text{Tr}[A^{\dagger}B]/d$ where $d$ is the Hilbert space dimension. Finally, the Krylov complexity is defined as
\begin{eqnarray}
    C_K(t) = \sum_{n} n|\phi_n(t)|^2.
\end{eqnarray}

\noindent 

For the very large dimensional case, if the Krylov space dimension is $K$, then the Krylov complexity for many-body chaotic systems or quantum circuits saturates at a value $K/2$ and saturates at time $K-1$.
In the operator case, the Krylov dimension $K$ is bounded by $1\leq K \leq d^2-d+1$. But for many-body chaotic systems, the Krylov space dimension is generally almost equal to $K \approx d^2-d+1$.

Since in our case, we consider three types of random quantum circuits (RUC, MLORC, MFORC), where the evolution of an initial state occurs due to repeated application of circuit layers at discrete times and there is no notion of Hamiltonian here, the above-mentioned method cannot be directly implemented here. We follow the prescription of ~\cite{Aranya@2024, Suchsland@2025} to compute the Krylov complexity in RU circuits. For MLORC and MFORC, we use a modified version that we shall discuss later.

\subsection*{Krylov complexity in circuits}

Let us take $\rho_0$ to be our initial state. We make it normalized (using the inner product definition mentioned before) and traceless. After the first layer of unitary $U$ (constructed as said before), our state becomes $\rho_1 = U\rho_0 U^{\dagger}$. In this way, we get the discrete time evolved set of states $\{\rho_0,\ \rho_1,\ \rho_2, \cdots\}$. We take the first Krylov basis operator $|K_0) = |\rho_0).$ Then we iteratively orthonormalize $|\rho_n)$ with all previous Krylov basis operators $\{ |K_0),  |K_1), ..., |K_{n-1})\}$ to get $|K_n)$. One thing to note here is that even though in the operator case the Krylov dimension $K$ is bounded by $1\leq K \leq d^2-d+1$, in our case the bound is given by $1\leq K\leq d^2-1$. This happens since in our case we apply different random unitaries in each layer, in contrast to the Hamiltonian picture, where $\mathcal{L}^n \rho_0$ means taking commutator $n$ times with the same Hamiltonian $H$. Applying different unitaries in each layer increases the Krylov dimension. Since the dimension of the operator space is $d^2$ (where the dimension of the underlying Hilbert space is $d$), and we are considering a traceless initial operator, the upper bound in our case becomes $d^2 -1$. If we use the same unitary $U$ on each layer, then we get to see the bound $d^2-d+1$.

\noindent For open quantum circuits, we follow the above-mentioned procedure. However, here, after each layer of evolution, we renormalize the state $\rho_n$ since, in the case of MLORC or MFORC, the evolution does not preserve the norm according to the innner product definition used. So, after renormalizing each state, we find the orthonormal set of states $\{\rho_0,\ \rho_1,\ \rho_2, \cdots\}$, using which, like before, we obtain the Krylov complexity measure $C_K(t)$.

\begin{figure}[h!]
    \centering
    \includegraphics[width=0.9\linewidth]{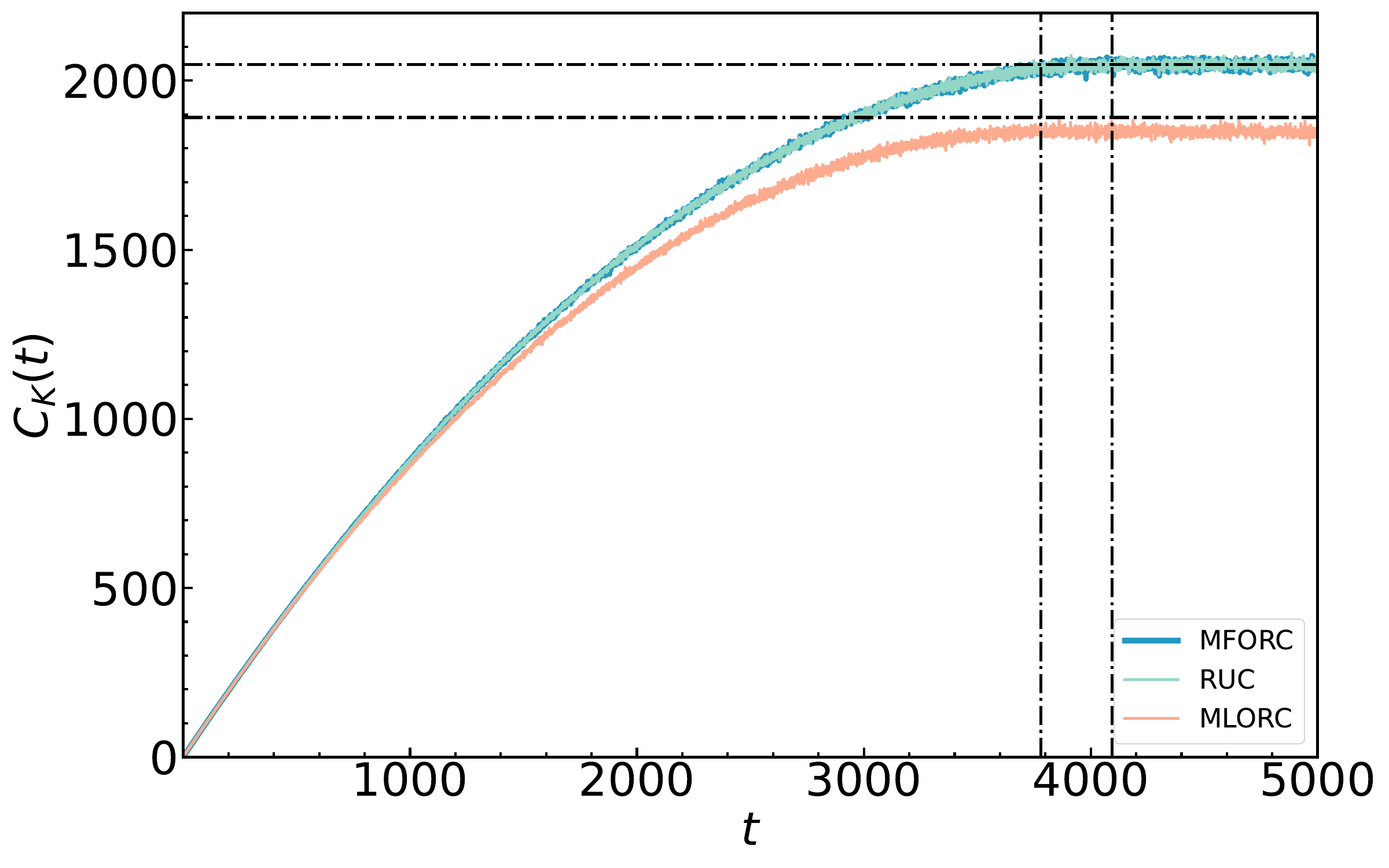}
    \caption{{\bf Krylov complexity growth in time.} Here, we have plotted the Krylov complexity of RUC, MLORC and MFORC for $N_s=6$ denoted by green, red, and blue smooth curves. 
    As shown in the plot for RUC and MFORC, the Krylov complexity saturates at a value close to $K/2=2047.5$, occurring at a time approximately equal to $K-1=4094$.
In contrast, for the MLORC circuit, we find that the Krylov dimension is reduced to $K = 3782$. Accordingly, the saturation value is slightly smaller than $K/2 = 1891$. For reference, we have indicated the lines $t = 3781$, $t = 4094$, $C_K(t) = 2047.5$, and $C_K(t) = 1891$ in the plot. The results are averaged over 10 realizations.}
    \label{fig:placeholder}
\end{figure}

 In our case, the Krylov space dimension of RUC and open random circuits with memory the bound is touched exactly, suggesting that these two circuits are maximally chaotic. However, for the memoryless open quantum circuit (MLORC), the dimension of the Krylov space is less than the bound. For the system size $N_s=6$, the bound on $K$ is  $4095$, which is exactly the dimension of Krylov space for RUC and MFORC circuits, but for MLORC it is much lower and given by $3782$. Corresponding results are shown in Fig.~\ref{fig:placeholder}.

To summarize the observations discussed above, ranging from quantum resource generation to dynamical complexity across the different classes of circuits considered in this work, we provide a schematic overview in Fig.~\ref{Schematic1}.

\section{Conclusion}\label{6s}

In this work, we have systematically investigated the generation of quantum resources and the emergence of dynamical complexity in random quantum circuits, extending the analysis beyond closed unitary dynamics to more realistic open-system settings involving system–environment interactions both with and without memory. 
From the perspective of quantum state preparation complexity, we analyzed the evolution of entanglement, mutual information, and non-stabilizerness across random unitary circuits (RUC), memoryless open random circuits (MLORC), and memoryful open random circuits (MFORC). We find that while RUC and MFORC exhibit sustained growth and saturation of entanglement and mutual information, MLORC shows a qualitatively distinct behavior where entanglement rapidly decays to zero due to strong environmental decoherence. Despite this suppression, non-stabilizerness remains finite even in MLORC, demonstrating that separable yet non-stabilizer states can still be generated.

The dynamical complexity, quantified through Krylov complexity, further reveals a clear distinction between memoryless and memoryful dynamics. Both RUC and MFORC exhibit strong chaotic behavior, with the Krylov-space dimension approaching its maximal bound, whereas MLORC shows a pronounced reduction in Krylov complexity, indicating suppressed operator spreading in the presence of memoryless environmental interactions.
To further probe the statistical nature of the generated state, we analyzed quantum state $k$-design. Our results showed that, although all circuits exhibit signatures of randomness at low moments, none fully approximate higher-order Haar randomness within accessible timescales. Notably, within the system sizes and timescales accessible in our simulations, 
memory effects enhance pseudorandomness, with MFORC outperforming MLORC and RUC behavior for $k=1$.

Overall, our results provide a unified framework for comparing different classes of random quantum circuits in terms of resource generation, dynamical complexity, and emergent randomness. More importantly, they demonstrate that distinct notions of quantum complexity respond qualitatively differently to environmental memory. While memoryless interactions suppress entanglement growth, operator spreading, and higher-order randomness generation, memoryful dynamics can preserve quantum correlations, sustain dynamical complexity, and enhance low-order pseudorandomness, often closely mimicking closed-system behavior. Although our analysis is restricted to finite system sizes, the observed trends already reveal the nontrivial role of environmental memory in shaping quantum complexity, making this framework relevant for benchmarking and controlling dynamics in near-term quantum devices. As discussed in the previous sections, future work will explore the individual roles of memory and environmental correlations in entanglement preservation, as well as the effect of varying the system–environment coupling strength through an explicit parameter.

 \section{Acknowledgment} S. S. R.  thanks 
Aranya Bhattacharya for useful discussion on Krylov complexity and  acknowledges financial support from the Faculty Research Scheme, IIT (ISM) Dhanbad, India, under Project
No. FRS/2024/PHYSICS/MISC0110, and from the Anusandhan National Research Foundation (ANRF), Government of India, under Grants No. ANRF/ARG/2025/004617/PS and
No. ANRF/ECRG/2025/002793/PMS. U.S. acknowledges support from ANRF, Government of India, under Grant No. ANRF/ARG/2025/004617/PS.
\appendix
\section{Growth of entanglement across different bipartitions in MLORC and MFORC}\label{app_1}

\begin{figure}[h!]
    \centering
    \includegraphics[width=1.0\linewidth]{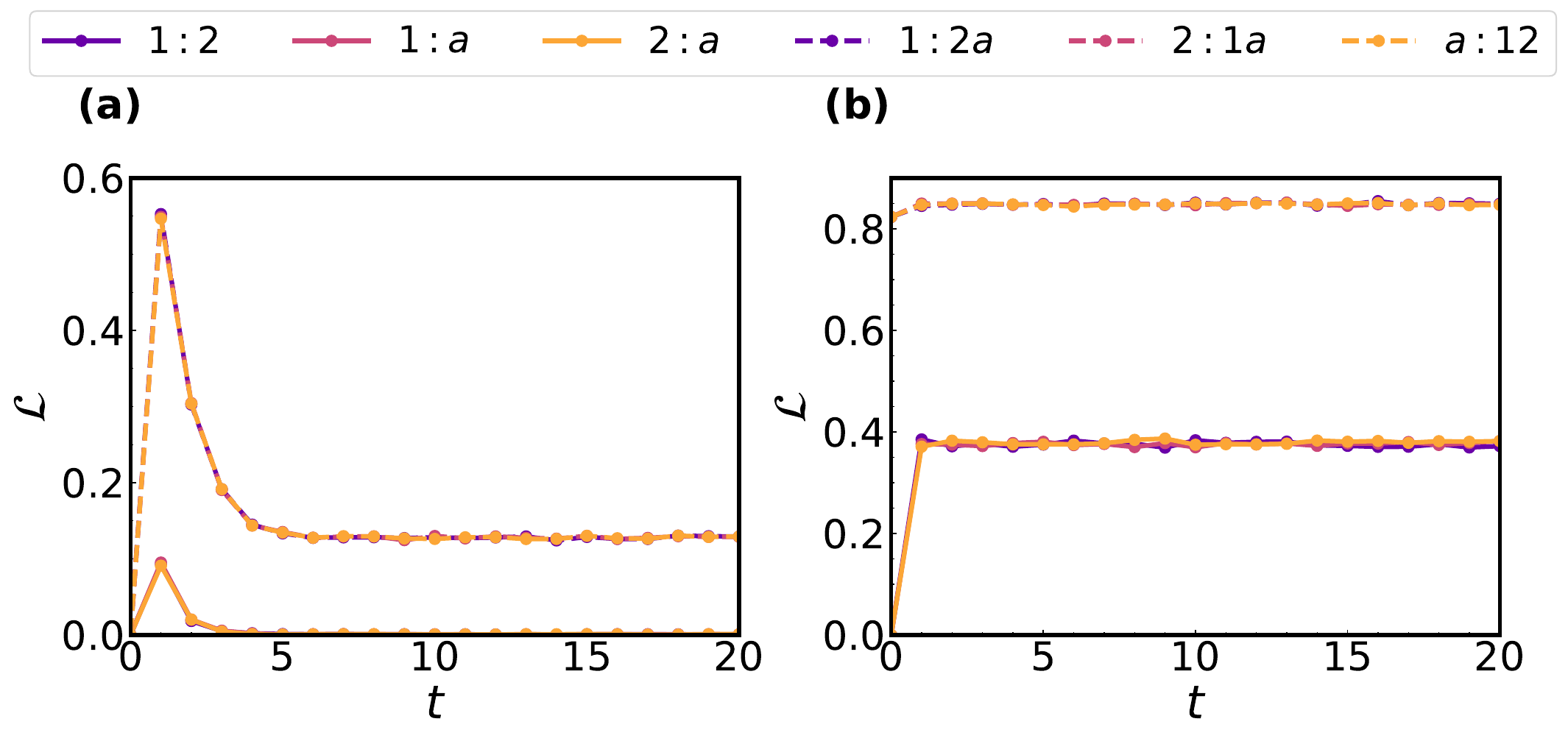}
  \caption{\textbf{Growth of entanglement across different bipartitions for MLORC and MFORC circuits.} Panels (a) and (b) depict the time evolution of the logarithmic negativity $\mathcal{L}$ as a function of time $t$ for different bipartitions involving the first two qubits and their associated auxiliary in a four-qubit memoryless open random circuit and an open random circuit with memory, respectively. Here, the labels $1$ and $2$ denote the system qubits on which the random circuit acts, while ``$a$'' represents the auxiliary qubit in both cases.
In both panels, the entanglement across the bipartitions $1:2a$, $2:1a$, and $a:12$ is shown by violet, red, and yellow dashed curves with markers, respectively. In contrast, the pairwise entanglement between $1:2$, $1:a$, and $2:a$ is represented by violet, red, and yellow solid curves with markers, where the complementary qubit (``$a$'', ``$2$'', and ``$1$'', respectively) is traced out in each case.
The qualitative differences between multipartite and pairwise entanglement reveal distinct entanglement structures in the two circuit architectures. Specifically, the memoryless open random circuit exhibits GHZ-like entanglement, characterized by strong global correlations accompanied by a suppression of pairwise entanglement. In contrast, the open random circuit with memory displays W-class–like behavior, wherein pairwise entanglement remains distributed among the subsystems.
}
    \label{af2x}
\end{figure}
Here, we provide a more detailed characterization of the structure of the evolved state once the entanglement saturates after repeated applications of the MLORC and MFORC circuits. Toward this goal, we consider a 4-qubit system and consider the first and second qubits with their corresponding auxiliary. Here, the labels ``$1$'' and ``$2$'' denote the first and second qubits of the system, respectively, while ``a'' denotes the auxiliary qubit. In Fig.~\ref{af2x}(a) we plot the logarithmic negativity along the vertical axis and time along the horizontal axis corresponding to MLORC. We examine the evolution of entanglement across different bipartitions of the system.  

The entanglement across the bipartitions $1:2a$, $2:1a$, and $a:12$ is shown by the violet, red, and yellow dashed curves with markers, respectively. We observe that the entanglement in all three bipartitions saturates to the same value, approximately $0.13$ (within numerical precision). At the same time, the pairwise entanglement between $1:2$, $1:a$, and $2:a$ is represented by the violet, red, and yellow solid curves with markers, where the qubits ``a'', ``$2$'', and ``$1$'' are traced out, respectively. Although the entanglement across any bipartition saturates at a finite, non-zero value, the pairwise entanglement vanishes in the long-time limit when one qubit is traced out. This behavior indicates that, at long times, the evolved state effectively exhibits generalized GHZ-state like entanglement.

A similar analysis for the MFORC circuit is shown in Fig.~\ref{af2x}(b), where the entanglement across the same bipartitions is represented using the same color and line styles. 
In particular, the pairwise entanglement between the bipartitions $1:2$, $1:a$, and $2:a$ become non-zero. However, the entanglement between the bipartitions $1:2a$, $2:1a$, and $a:12$ remains equal but larger than the pairwise entanglement. This pattern resembles the characteristic entanglement distribution of W state. Therefore, although the initial state in MFORC is chosen randomly, at later times the entanglement structure evolves toward a W-state–like state. We further check that this intriguing feature holds for any two qubit pair and their corresponding random auxiliary for both MLORC and MFORC.

\bibliography{references}{}

\end{document}